\documentclass[journal]{IEEEtran}

\usepackage{cite}
\usepackage{autobreak}
\usepackage{amsmath,amssymb,amsfonts}
\usepackage{graphicx}
\usepackage{textcomp}
\usepackage{xcolor}

\usepackage{hyperref}
\usepackage{amsmath}
\usepackage{autobreak}
\usepackage{amsfonts}
\usepackage{bm}
\usepackage{bbm}
\usepackage{algorithm}  
\usepackage{algorithmicx}  
\usepackage{algpseudocode}
\usepackage{stfloats}
\usepackage{cuted}
\usepackage{color}
\usepackage{subfigure}
\usepackage{cite}
\usepackage{booktabs} 
\definecolor{red}{RGB}{0,0,0}
\definecolor{bl}{RGB}{30,200,230}

\newcommand{\tr}{\text{Tr}}

\newcommand{\hk}{\mathbf{h}_k}

\newcommand{\red}[1]{{\color{red}#1}}

\newcommand{\SINR}{\text{SINR}}
\newcommand{\SCNR}{\text{SCNR}}
\newcommand{\SNR}{\text{SNR}}
\newcommand{\CRB}{Cram\'er-Rao }
\makeatletter
\renewcommand{\maketag@@@}[1]{\hbox{\m@th\normalsize\normalfont#1}}%
\makeatother

\ifCLASSINFOpdf
\else
\fi

\begin{document}

\title{Movable Antenna Enabled Integrated Sensing \\ and Communication}

	\author{Wanting Lyu,~Songjie Yang,~Yue Xiu,~Zhongpei Zhang,~\IEEEmembership{Member,~IEEE},\\Chadi Assi,~\IEEEmembership{Fellow,~IEEE}, and Chau Yuen,~\IEEEmembership{Fellow,~IEEE} 
		
		\thanks{
        This work was supported in part by the Natural Science Foundation of
        Shenzhen City under Grant JCYJ20210324140002008; and in part by key R\&D program of Zhejiang under Grant No.Z24JX007 (Corresponding author: Zhongpei Zhang).
        
        Wanting Lyu, Songjie Yang, and Yue Xiu are with National Key Laboratory of Wireless Communications, University of Electronic Science and Technology of China, Chengdu 611731, China (E-mail: lyuwanting@yeah.net; yangsongjie@std.uestc.edu.cn; xiuyue12345678@163.com).

        Zhongpei Zhang is with Shenzhen institute for Advanced Study, University of Electronic Science and Technology of China, Shenzhen 518110, China and Donghai Laboratory, Zhoushan 316021, China (zhangzp@uestc.edu.cn).
			
		Chadi Assi is with Concordia University, Montreal, Quebec, H3G 1M8, Canada (email:assi@ciise.concordia.ca).
			
		Chau Yuen is with the School of Electrical and Electronics Engineering, Nanyang Technological University, 639798 Singapore (E-mail: chau.yuen@ntu.edu.sg).
        
        This paper has been accepted by IEEE Transactions on Wireless Communications. The conference version has been accepted by 2024 IEEE 24th International Conference on Communication Technology (available: https://arxiv.org/abs/2405.10507).}

		}

\maketitle

\begin{abstract}

In this paper, we investigate a novel integrated sensing and communication (ISAC) system aided by movable antennas (MAs). A bistatic radar system, in which the base station (BS) is configured with MAs, is integrated into a multi-user multiple-input-single-output (MU-MISO) system. Flexible beamforming is studied by jointly optimizing the antenna coefficients and the antenna positions. Compared to conventional fixed-position antennas (FPAs), MAs provide a new degree of freedom (DoF) in beamforming \red{to reconfigure} the field response, and further improve the received signal quality for both wireless communication and sensing. We propose a communication rate and sensing mutual information (MI) maximization problem by flexible beamforming optimization. The complex fractional objective function with logarithms are first transformed with the fractional programming (FP) framework. Then, we propose an efficient algorithm to address the non-convex problem with coupled variables by alternatively solving four sub-problems. We derive the closed-form expression to update the antenna coefficients by Karush-Kuhn-Tucker (KKT) conditions. To improve the direct gradient ascent (DGA) scheme in updating the positions of the antennas, a 3-stage search-based projected GA (SPGA) method is proposed. Simulation results show that MAs significantly enhance the overall performance of the ISAC system, achieving 59.8\% performance gain compared to conventional ISAC system enabled by FPAs. Meanwhile, the proposed SPGA-based method has remarkable performance improvement compared the DGA method in antenna position optimization. 

\end{abstract}

\begin{IEEEkeywords}
	Movable antenna, integrated sensing and communication, dual functional radar and communication, beamforming.
\end{IEEEkeywords}

\IEEEpeerreviewmaketitle

\section{Introduction}
\IEEEPARstart{I}{ntegrated} sensing and communication (ISAC) has become a promising technology in the next generation mobile networks \cite{ISAC1,ISAC2,JSAC_ISAC}. By sharing the hardware platform and signal processing modules, ISAC enables efficient resource utilization for communication and radar sensing, significantly enhancing spectral efficiency and hardware costs \cite{8999605,ISAC3}. Through strategic collaboration involving the appropriate reuse of resources and information, ISAC enhances both localization and communication performance \cite{JSTSP_ISAC}. Attributable to these advantages, ISAC have shown great potential in a variety of applications, including vehicle-to-everything (V2X), industrial internet of things (IIoT), and environment monitoring \cite{ISAC1}. Moreover, by integrating emerging wireless technologies, such as reconfigurable intelligent surface (RIS) \cite{CRB1,RIS_ISAC}, unmanned aerial vehicle (UAV) \cite{UAV_ISAC}, and non-orthogonal multiple access (NOMA) \cite{BP1,NOMA2}, ISAC has attracted significant attention of both academia and industrial.

In conventional multiple-in-multiple-out (MIMO) systems, fixed-position antennas (FPAs) are configured and widely employed  both for sensing and communication. Although the waveform or precoding matrix is programmable to improve sensing accuracy or channel capacity, the wireless channel remains uncontrollable within a certain transmit and receive region. Lots of research has been dedicated to explore antenna position optimization technologies to enhance the spatial diversity of the wireless channels. As one of the solutions, antenna selection enables dynamically selecting a subset of antennas in a dense array to improve the channel capacity \cite{AS}. Array synthesis is another technique exploiting flexible array structure to pursue degrees of freedom (DoFs) for archiving some beampattern indicators, such as given a desired beampattern, a minimum sidelobe power, and the array sparsity, for optimizing antenna excitation coefficients, antenna positions \cite{AntSyn2, AntSyn}.

Despite considerable performance improvement brought by these two techniques, the feasible positions of the antennas are discrete and pre-determined in antenna selection, which cannot fully utilize the entire array range. Furthermore, array synthesis focuses more on radiation patterns rather than the physical channel in wireless communications. Recently, movable antenna (MA), similar as the concept of fluid antenna system (FAS), has been investigated to provide additional DoFs for beamforming \cite{MA-mag,FA}. Hence, to fully explore the spatial diversity, MA/FA is emerging to flexibly reconfigure the wireless channels. This is achieved by strategically designing the positions of antennas in a continuous region to modify the field response \cite{MA1}. With great potential in wireless communication, MA and FAS have attracted attention for performance analysis and optimization design.

\subsection{Related Works}

Prior studies on ISAC have obtained insightful results in waveform design with conventional MIMO. The primary goal of ISAC is to enhance both communication capacity and sensing ability \cite{ISAC3_twc}. Different metrics have been employed in the literature to evaluate sensing performance. One is beampattern gain maximization at the target angles, which ensures high beampattern gain for target sensing \cite{BP1,NOMA2}. Authors in \cite{BP2} proposed a beampattern matching problem, aimed at minimize the matching error between idealized beampattern and designed one. Considering estimation accuracy, minimizing the \CRB Bound (CRB) has become another important objective in ISAC beamforming design \cite{CRB1,CCC,CRB2,CRB3,ISAC2_jsac}. In more practical cluttered environments, the radar signal-to-clutter-plus-noise-ratio (SCNR) has been exploited in \cite{PMN}. Optimizing SCNR is beneficial for enhancing target echoes and suppressing clutter echoes. \red{With similar concept as communication MI indicating achievable data rate, sensing MI quantifies the MI between the received echo signal at the sensing receiver and the estimated target parameter, given the knowledge of the probing signal \cite{SMI_CommMag}. Sensing MI has been widely investigated as the metric to measure target sensing performance in \cite{SMI_ISIT,SMI2,SMI3,ISAC1_twc}.}

\red{Balancing the demands of sensing and communication inevitably involves trade-offs in performance and resource allocation, and employing varied metrics for sensing tasks can lead to differing outcomes.} In \cite{BP2,ISAC4,NF_ISAC_CRB}, sensing-focused systems were investigated by proposing an optimization problem that prioritizes sensing performance as the objective function, with communication performance serving as the constraint. Authors in \cite{BP2} formulated a radar beampattern matching problem aimed at minimizing the matching error between the optimized beampattern and the idealized value, while meeting the minimum requirements for communication signal-to-interference-plus-noise (SINR). Authors in \cite{ISAC4} studied a simultaneously transmitting and reflecting surface (STARS) enabled ISAC system, where the CRB for estimating target angles was minimized, subject to a minimum SINR constraint for communication users. \red{In \cite{NF_ISAC_CRB}, the authors proposed two algorithms for downlink and uplink sensing and communication in a near-field scenario, respectively. To improve the positioning accuracy, CRB for both angle and distance was derived as the sensing metric and minimized.} Conversely, in \cite{RIS_ISAC_TVT}, a communication-focused ISAC system aided by RIS was explored. The sum of communication rate of the multiple users was regarded as the objective function, with constraint on the minimum tolerable radar beampattern matching error. In practical engineering, the priority of the two functions may vary. Consequently, a more flexible framework is commonly used to maximize the combined performance metrics of communication and sensing in the objective function, with an adjustable weighting factor. In \cite{LF_18TSP}, the shared waveform between signal transmission and radar sensing was optimized to minimize the weighted sum of multi-user interference and radar waveform matching error. Authors in \cite{ZHL_22CL} proposed a problem to maximize the weighted combination of the sensing MI and the communication data rate by jointly optimizing the beamforming matrix and RIS phase shifts. \red{A more comprehensive research was conducted by the authors of \cite{SMI2}, where two types of waveforms were designed respectively using CRB and sensing MI as sensing performance metrics. The trade-off between CRB/MI and the communication sum rate was explored , providing comparisons between waveforms designed with different performance criteria.} \red{ Nevertheless, the aforementioned works are all based on FPAs, where the channels for both communication and sensing are not adjustable. This configuration may restrict the full exploration of available DoFs in the continuous spatial domain, thereby resulting performance limitation in spatial diversity and multiplexing multi-user service and target sensing.}

\red{Several studies have investigated the integration of MA/FAS to enhance communication capacity \cite{MASISO, MA2-MIMO,MA-MU2,MA-MU4,FA-MISO,MA-DOG,MAMU6,FAS,MA_null,ISAC_smart_propagation}.} Specifically, the performance of the MA enabled wireless communication system was analyzed in \cite{MASISO}. In this study, both the transmitter and the receiver were equipped with a single MA. Closed-form expressions for the channel gains were derived for both deterministic and stochastic channel environments, demonstrating the effectiveness of MA in enhancing channel capacity. In \cite{MA2-MIMO}, using MA in multiple-input multiple-output (MIMO) was analyzed, with MAs utilized at both transmitter and receiver. Results showed that the capacity was significantly increased by strategically designing antenna positions. Multi-user multiple-input single-output (MU-MISO) systems, with the base station (BS) configured with MAs, were studied in \cite{MA-MU2,MA-MU4}. In \cite{MA-MU2}, a particle swarm optimization (PSO)-based algorithm was employed to determine the optimal positions for the antennas. In \cite{MA-MU4}, fractional programming (FP) and gradient ascent (GA) methods were employed to optimize the positions of antennas, thereby enhancing the sum rate of users. Additionally, the zero-forcing (ZF) method was utilized to design the precoding matrix. Equipping single fluid antenna at the users, the authors of \cite{FA-MISO} investigated a power minimization problem, where the FA for each user is empowered to move at a 2-dimensional region. Flexible precoding with MAs was proposed by the authors of \cite{MA-DOG}, which not only adjusted the antenna coefficients (corresponding to traditional precoding) but also optimized element positions under a sparse optimization framework, improving traditional precoding schemes. In \cite{MAMU6}, MA was utilized to enhance physical layer security (PLS). A projected GA-based algorithm was proposed to design the antenna positions in a linear array. In \cite{FAS}, the outage probability of an FA system was analyzed, where $M$ best ports were activated among totally $K$ ports at the receiver for signal combining. The authors in \cite{MA_null} proposed a null steering scheme enabled by MA, providing valuable insights for interference cancellation. ZF precoding was applied to implement null steering in undesired directions, while the antenna positions were further optimized to maximize the receive signal power in the desired direction.

In addition to wireless communication, the authors of \cite{MA_sensing} studied MAs in enhancing wireless sensing performance. The CRB for angle of arrival (AoA) estimation was minimized by designing the antenna positions in linear, planar, and circular arrays. The results indicated that the MA array significantly improved the accuracy of target AoA estimation. \red{In \cite{MA_RIS_ISAC}, a RIS is adopted to assist MA enabled ISAC system. Significant performance improvement was obtained with the assistance of both RIS and MA.}

\subsection{Motivation and Contributions}

Based on the above discussion, \red{MAs theoretically demonstrate significant potential in both communication and sensing by flexibly configuring wireless channels. Thus, application of MAs in ISAC systems is crucial for enhancing spectrum utilization, increasing channel capacity, and improving mutual interference suppression. However, research on MA-aided ISAC is still in its infancy.} To bridge this gap, this paper studies flexible beamforming in an ISAC system where the BS is equipped with an MA array, aiming to simultaneously enhance both communication and sensing performance. \red{Despite the benefits of MAs in both communication and sensing systems, optimizing the antenna positions to effectively enhance the performance of both tasks remains a challenging issue. The main difficulty arises from the complex expression of the wireless channel with respect to the positions of the antennas, causing non-convex expressions involving coupling variables.} To the best of the authors' knowledge, this is among the first works to explore MA-based ISAC. The main contributions of this paper are summarized as follows.

\begin{itemize}
	\item 
	We study a bistatic ISAC system in which a dual functional radar and communication (DFRC) BS serves multiple users and sense a target. \red{Aiming to further explore the spatial DoFs and diversity gains, we deploy an MA array at the BS as the transmitter, and we introduce flexible beamforming that simultaneously optimizes both the antenna coefficients (transmit beamforming) matrix and the antenna positions.}
	
	\item
	For practical considerations, we study a cluttered environment as a source of sensing interference. Thus, the communication rate and sensing MI are derived as the performance metrics. We then propose a problem to maximize the sum of communication rate and sensing MI by optimizing both beamforming and the positions of the transmit antennas. A weighting factor is adopted to control the priority of the two functions.
	
	\item
	Fractional programming (FP) framework is employed to transform the fractional objective function with logarithms. Utilizing alternating optimization (AO) method, the problem is decomposed into four sub-problems to handle the coupling variables. Specifically, we utilize Karush–Kuhn–Tucker (KKT) conditions and direct gradient ascent (DGA) method to solve beamforming and antenna positions, respectively, referred to as DGA-based flexible beamforming. Regarding the significance of initial points in DGA and aiming to fully exploit the feasible moving region, we propose a 3-stage search-based projected gradient ascent (SPGA) method to replace DGA, leading to significant performance improvement. 
	
	\item
	We perform numerical simulations to evaluate the performance of MA in ISAC system with our proposed algorithm. \red{Simulation results confirm that adopting MA significantly enhances both sensing and communication performance with our SPGA-based algorithm.} In addition to MI, we obtain the beampattern diagram and the CRB to verify the effectiveness of MA on sensing performance.
\end{itemize}

\subsection{Organizations and Notations}

The rest of this paper is organized as follows. In Section II, the system model of the MA enabled ISAC system is described, and a flexible beamforming problem is proposed. In Section III, we propose an efficient algorithm to solve the complex problem by alternatively optimizing the beamforming matrix and the antenna positions. Finally, the simulation results and the conclusion are given in Section IV and Section V, respectively.

In this paper, vectors and matrices are denoted by boldface lower and upper-case letters, respectively. Conjugate, transpose and conjugate transpose of $\mathbf x$ are denoted by $\mathbf x^*$, $\mathbf x^T$ and $\mathbf x^H$, respectively. $|x|$ and $||\mathbf x||$ denote the modulus of the scalar $x$ and the $l_2$-norm of vector $\mathbf x$.  $\mathbb E\{\cdot\}$ is the expectation operator, $\mathcal Re\{\cdot\}$ and $\mathcal Im\{\cdot\}$ are the real and imaginary part of the parameter, respectively. $\tr(\cdot)$ denotes the trace of the parameter. $\mathbf I_N$ is the $N\times N$ identity matrix, and $\mathbf X^{\dagger}$ represents the pseudo inverse of matrix $\mathbf X$.   

\section{System Model and Problem Formulation}

\begin{figure}[t]
	\centering
	\includegraphics[width=0.95\linewidth]{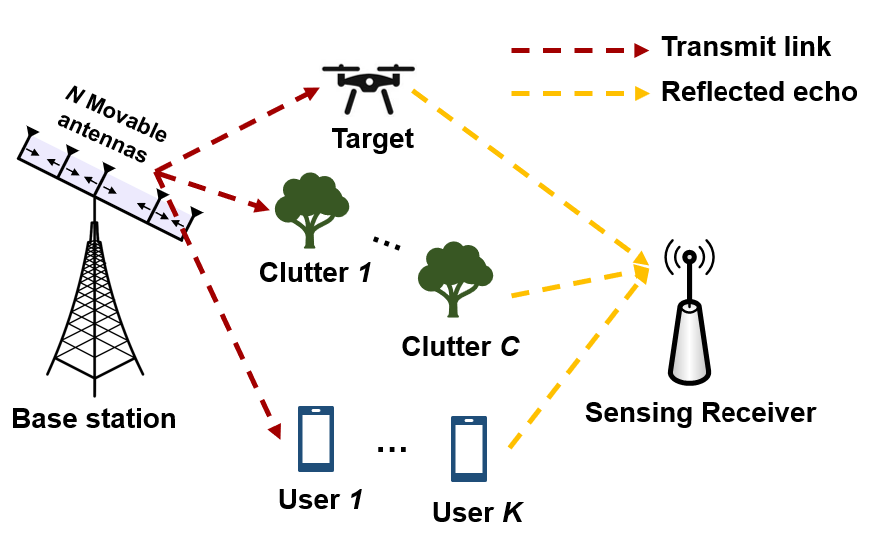}
	\caption{System model of the MA-ISAC system.}
	\label{system model}
\end{figure}

Consider a dual functional radar and communication (DFRC) BS serving $K$ users and sensing one target as shown in Fig. \ref{system model}. Instead of monostatic sensing system, where the transmitter (Tx) and the receiver (Rx) are co-located, we study a bistatic system with separated Tx and Rx to avoid challenging issues in full-duplex mode. Thus, self-interference can be effectively avoided and ignored assuming the Tx and Rx are sufficiently separated \cite{PMN}.  To study a more practical scenario, we assume there are $C$ clutters as the interference for sensing. In this paper, we consider the transmit BS is configured with a linear array with $N$ MAs. Each user, as well as the sensing receiver (SR) are equipped with single FPA.

\subsection{Channel Model}

Consider the far-field assumption, where the dimension of the Tx array is sufficiently small compared to the distance of wave propagation. Consequently, for each channel path component, all antenna elements within the same Tx array experience the same angle of departure and the same amplitude of the path gain. Hence, the array response is based on the different phase of the complex path coefficients among different antenna elements \cite{MA2-MIMO}. 

According to \cite{MASISO}, the MAs are connected to the radio frequency chains through flexible wires like coaxial cables. This allows the positions of the MAs to be dynamically adjusted using drive components such as stepper motors. In this paper, we consider the 1-dimension array for simplicity. We denote the positions of the transmit MAs as $\mathbf x = [x_1, x_2 \cdots, x_N]^T$. \footnote{Note that when uniform linear array with FPAs is adopted, $x_n$ can be set as $x_n = (n-1)d_x$, where $d_x$ is the distance of adjacent elements.} Employing the classic multi-path model for communication users. The field response vector (FRV) of the $l$-th propagation path of the $k^{th}$ user can be expressed as
\begin{equation}
	\mathbf a_{k,l}(\mathbf x) = \left[e^{j\frac{2\pi}{\lambda}x_1\cos\theta_{k,l}}, \cdots, e^{j\frac{2\pi}{\lambda}x_N\cos\theta_{k,l}} \right]^T \in \mathbb C^{N\times 1}, \label{array_linear}
\end{equation} 
where $\theta_{k,l}$ denotes the angle of direction (AoD) of the $l$-th path of user $k$. Assume that the channel from the BS to each user experiences $L_p$ paths. Hence, the communication channel between the Tx and user $k$ can be expressed as
\begin{equation}
	\mathbf h_k(\mathbf x) = \sqrt{\frac{N}{L_p}}\sum_{l=1}^{L_p} \rho_{k,l}\mathbf a_{k,l}(\mathbf x),
\end{equation}
where $\rho_{k,l}$ is the complex channel gain for the $l$-th path of user $k$.

Without loss of generality, single path model is applied for sensing, since multi-hop reflected signals may experience severe path loss. Thus, we assume that only the line-of-sight (LoS) path from the BS to the target/clutters, and from the target/clutters to the SR can be successfully detected. Hence, the FRV from the BS to the sensing target/clutters can be written as
\begin{equation}
	\mathbf a_\chi(\mathbf x) = \left[e^{j\frac{2\pi}{\lambda}x_1\cos\theta_\chi}, \cdots, e^{j\frac{2\pi}{\lambda}x_N\cos\theta_\chi} \right]^T \in \mathbb C^{N\times 1},
\end{equation} 
where $\chi = \{s,1,\cdots,C\}$ representing the target or the clutter.

\subsection{Communication and Sensing Signal Model}

We assume $K+1$ streams are transmitted as $\mathbf s = [s_1, \cdots, s_K, s_{K+1}]^T \in \mathbb C^{K_1\times 1}$, where $s_1,\cdots, s_K$ are for the $K$ communication users, respectively, and $s_{K+1}$ is dedicated for sensing. Consider that the SR knows the transmit signal $\mathbf s$, and thus the total $K+1$ data streams can be utilized for sensing. Without loss of generality, we assume $s_1, \cdots, s_{K+1}$ are identically complex Gaussian distributed and mutually independent with zero mean and unit variance, satisfying $\mathbb E\{\mathbf s\mathbf s^H\} = \mathbf I$. \red{Assume that the BS perfectly knows the channel state information (CSI), where the CSI can be obtained by channel estimation methods for MA-aided systems in \cite{Channel_estimation_CL,Channel_estimation_TWC,Channel_estimation_WCL}. } Correspondingly, the transmit beamforming matrix can be written as
\begin{equation}
	\mathbf F = \underbrace{[\mathbf f_1, \cdots, \mathbf f_K}_\text{for communication and sensing},\underbrace{ \mathbf f_{K+1}]}_\text{dedicated for sensing} \in \mathbb C^{N\times{(K+1)}}.
\end{equation}

Based on the above discussions, the received signal at user $k$ can be expressed as
\begin{equation}
	y_k = \underbrace{\hk^H(\mathbf x)\mathbf f_k s_k}_\text{interested signal} + \underbrace{\sum_{j=1,j\neq k}^{K+1} \hk^H(\mathbf x)\mathbf f_j s_j}_\text{multi-user and target interference} + n_k,
\end{equation}
where $n_k \sim \mathcal{CN}(0,\sigma_k^2)$ is the additive white Gaussian noise (AWGN). $\mathbf h_k(\mathbf x)$ denotes the channel between the Tx and user $k$ that is related to antenna position $\mathbf x$. Hence, the received data rate of user $k$ can be obtained as
\begin{equation}
	R_k = \log_2(1+\text{SINR}_k),
\end{equation}
where 
\begin{equation}
	\SINR_k = \frac{\left|\hk^H(\mathbf x)\mathbf f_k\right|^2}{\sum_{j=1,j\neq k}^{K+1}\left|\hk^H(\mathbf x)\mathbf f_j\right|^2 +\sigma_k^2}.
\end{equation}

As for sensing, the transmitted sensing and communication signals are reflected by the sensing target and the clutters, then received by the SR. Without loss of generality, we assume point-like target and clutters, since the sizes of the target and the clutters are sufficiently small compared to the distances of the reflecting paths. Thus, the echo signal for sensing can be expressed as
\begin{equation}
	y_s = \underbrace{\alpha_s\mathbf a_s^H(\mathbf x) \mathbf{Fs}}_\text{interested echo by the target} + \underbrace{\sum_{c=1}^C \alpha_c\mathbf a_c^H(\mathbf x)\mathbf{Fs}}_\text{interference echo by clutters} + n_s,
\end{equation}
where $n_s \sim\mathcal{CN}(0,\sigma_s^2)$ denotes the AWGN for radar link. $\alpha_s$ and $\alpha_c$ are complex coefficients including the radar cross section (RCS) of the target/clutter $c$, and cascaded complex gains of the target/clutter $c$, respectively. Additionally, $\mathbf a_s(\mathbf x)$ and $\mathbf a_c(\mathbf x)$ denote the array response vectors between the BS and the target/clutter $c$, respectively. 

Thus, we can obtain the radar signal-to-clutter-plus-noise-ratio (SCNR) at the Rx as 
\begin{equation}
	\SCNR = \frac{\left\Vert\alpha_s\mathbf a_s^H(\mathbf x)\mathbf F\right\Vert^2}{\sum_{c=1}^C\left\Vert\alpha_c\mathbf a_c^H(\mathbf x)\mathbf F\right\Vert^2 + \sigma_s^2}. \label{exp_SCNR}
\end{equation}

While some studies use the radar SCNR as the sensing metric, to mathematically align with the logarithmic form of the communication rate, sensing MI is employed to evaluate sensing performance \cite{SMI3}. Since the SR has prior knowledge of the transmitted waveform for both communication and sensing functions for target detection and estimation, based on \cite{InformationTheory,MI_TAES,MI_ICT}, the conditional sensing MI between the radar array response and the received echo can be further derived as
\begin{equation}
	R_s = I(y_s;\alpha_s\mathbf a_s|\mathbf s_t) = \log_2(1+\SCNR),
\end{equation}
where $\mathbf s_t = \mathbf{Fs}$ denotes the transmitted waveform.

\subsection{Problem Formulation}
\begin{figure}[t]
	\centering
	\includegraphics[width=1\linewidth]{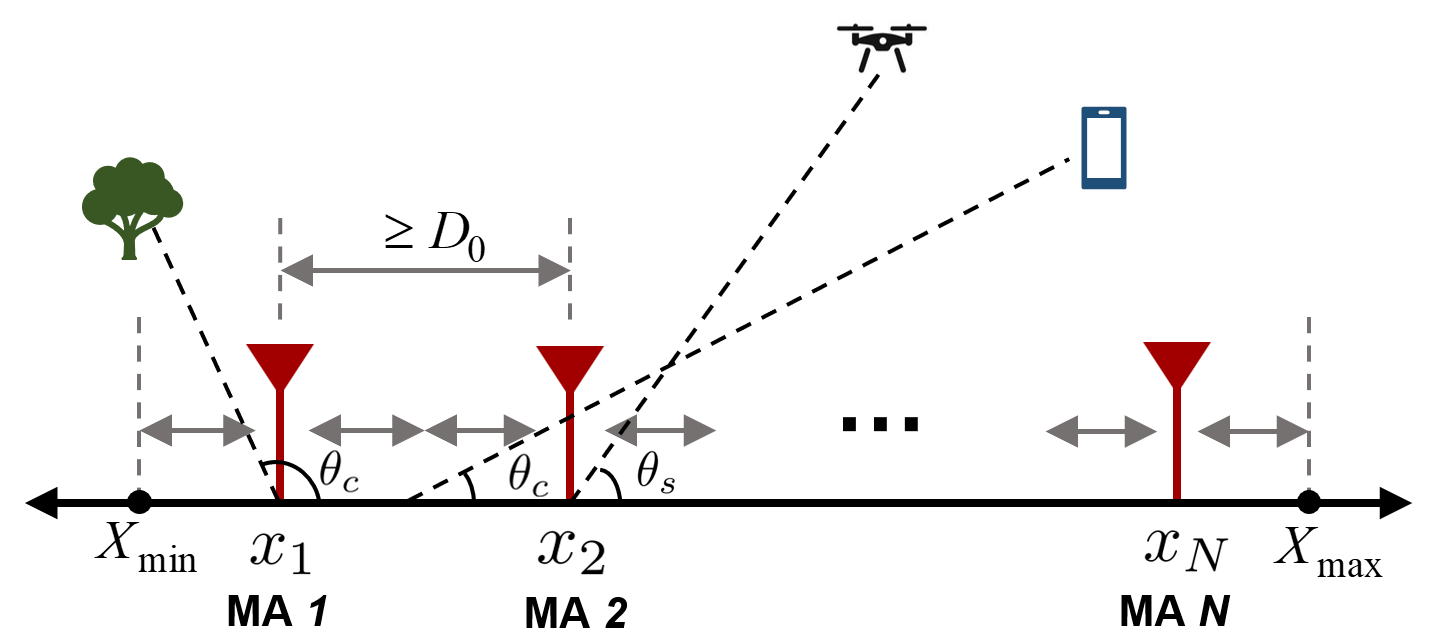}
	\caption{Movable antenna model for linear array.}
	\label{MAmodel}
\end{figure}

As mentioned before, the antenna elements at the BS are capable of moving along a given line segment as illustrated in Fig. \ref{MAmodel}. \footnote{\red{Please note that the original sequence of antenna elements may be altered after optimizing their positions. To fully leverage the flexibility of the beamforming gain offered by the MAs, no strict sequence constraints are imposed on the antenna elements during the optimization process.}} To mitigate the coupling effect between the elements in the MA array, it is important to ensure a sufficient separation distance between the adjacent elements:
\begin{equation}
	\red{\left\vert x_z - x_q \right\vert \ge D_0,\; z,q = 1,\dots,N, z\neq q,} \label{Cons_dis_linear}
\end{equation}
where $D_0$ denotes the minimum acceptable distance between two adjacent elements. Besides, the feasible region is predetermined as $\mathcal A = [X_\text{min},X_\text{max}]$.

We aim at maximizing the sum of communication rate and sensing MI. To control the priority of communication and sensing, we introduce weighting factors $\varpi_c$ and $\varpi_s$ for the two functions. Accordingly, the optimization problem can be formulated as
\begin{align}
(\text{P1}) \;\max_{\mathbf F,\mathbf x}\; & \mathcal G(\mathbf F,\mathbf x) = \varpi_c\sum_{k=1}^K R_k + \varpi_s R_s, \label{objfunc} \\
\text{s. t. } & \tr(\mathbf F^H \mathbf F) \le P_0, \label{Cons_pow} \tag{\ref{objfunc}a}\\
& \mathbf x \in \mathcal A, \label{Cons_ant_region} \tag{\ref{objfunc}b}  \\
& (\ref{Cons_dis_linear}), \notag
\end{align}
where $\varpi_c + \varpi_s = 1$. (\ref{Cons_pow}) denotes the transmit power budget constraint, $(\ref{Cons_ant_region})$ is for feasible moving region. However, this problem is non-convex with coupled variables. \red{The primary challenge in solving this problem stems from the non-convex fractions within the objective function, which combines both data rate and sensing MI. These fractions involve complex expression of antenna positions in the array response. Additionally, the non-convex constraints (\ref{Cons_ant_region}) and (\ref{Cons_dis_linear}), as well as the coupling of variables makes the problem more difficult to address. Moreover, the trade-off between communication and sensing needs to be investigated, due to the resource allocation to meet the different demands of the two functions. }

\red{It is worth mentioning that, for simplicity, we focus on a single target in this paper for simplicity, even though in practice, multiple sensing targets may interfere with each other. However, our proposed scheme can be readily extended to multi-target systems by treating the echoes from other targets as radar interference in the denominator of (\ref{exp_SCNR}). Due to the similar mathematical structure, the algorithm proposed here can be similarly adapted to address the multi-target problem.}

%

\section{Proposed SPGA-FP Based Approach for Flexible Beamforming}

In this section, we propose an efficient algorithm for the proposed flexible beamforming problem by optimizing the beamforming matrix and the antenna positions.

The objective function is multi-ratio with logarithms, and the numerators as well as the denominators are convex with quadratic forms with respect to $\mathbf F$, $\mathbf a_{k,l}(\mathbf x)$, $\mathbf a_s(\mathbf x)$, and $\mathbf a_c(\mathbf x)$. Based on the FP framework \cite{FP2}, we employ the Lagrangian dual transform and quadratic transform to equivalently transform $\mathcal G(\mathbf F,\mathbf x)$ into (\ref{objFP}) at the bottom of this page, where $\bm\mu = [\mu_1,\cdots,\mu_{K+1}]$, $\bm\xi^c = [\xi_1^c,\cdots,\xi_K^c]^T$ and $\bm\xi^s = [\xi_1^s,\cdots,\xi_{K+1}^s]^T$ denote auxiliary variables. The details of the transformation can be seen from \cite{FP2,PMN}.
\begin{figure*}[b]
	\hrule
\begin{equation}
	\begin{split}
	&\tilde{\mathcal G}(\mathbf F,\mathbf x,\bm\mu,\bm\xi^c,\bm\xi^s) = \varpi_c\sum_{k=1}^{K}\ln(1+\mu_k)+\varpi_s\ln(1+\mu_{K+1})  - \varpi_c\sum_{k=1}^{K} \mu_k  - \varpi_s\mu_{K+1} \\
	&\qquad+ \varpi_c\sum_{k=1}^K\Bigg(2\sqrt{1+\mu_k}\mathcal Re\left\{ \xi_k^c\hk^H(\mathbf x)\mathbf f_k\right\}  - \vert\xi_k^c\vert^2 \left(\sum_{j=1}^{K+1}\left\vert\hk^H(\mathbf x)\mathbf f_j\right\vert^2 +\sigma_k^2\right)\Bigg) \\
	& + \varpi_s \left( 2\sqrt{1+\mu_{K+1}}\mathcal Re\left\{ \alpha_s\mathbf a_s^H(\mathbf x)\mathbf F\bm\xi^s \right\} - \Vert\bm\xi^s\Vert^2\left( \sum_{c=1}^C\left\Vert\alpha_c\mathbf a_c^H(\mathbf x)\mathbf F\right\Vert^2 + \left\Vert \alpha_s\mathbf a_s^H(\mathbf x)\mathbf F\right\Vert^2 + \sigma_s^2\right) \right).
	\end{split} \label{objFP}
\end{equation}
\end{figure*}

Due to the mutually coupled variables, we propose an alternating optimization (AO)-based algorithm to solve the transformed problem iteratively. Specifically, in each iteration, we update each variable with fixed values of the other variables obtained from the last iteration. (P1) can be decomposed as four sub-problems as follows:
\begin{itemize}
	\item (SP.1) updating beamforming matrix $\mathbf F$:
	\begin{align}
		\max_\mathbf{F} \;&\tilde{\mathcal G}\left(\mathbf F|\mathbf x^{(t)},\bm\mu^{(t)},\bm\xi^{c(t)},\bm\xi^{s(t)}\right), \\
		\text{s. t. } &(\ref{Cons_pow}). \notag
	\end{align}

	\item (SP.2) updating antenna positions $\mathbf x$:
	\begin{align}
		\max_\mathbf{x} \;&\tilde{\mathcal G}\left(\mathbf x|\mathbf F^{(t)},\bm\mu^{(t)},\bm\xi^{c(t)},\bm\xi^{s(t)}\right), \\
		\text{s. t. } &(\ref{Cons_ant_region}), (\ref{Cons_dis_linear}). \notag
	\end{align}

	\item (SP.3) updating auxiliary variable $\bm\mu$:
	\begin{equation}
		\max_{\bm\mu} \;\tilde{\mathcal G}\left(\bm\mu|\mathbf F^{(t)}, \mathbf x^{(t)},\bm\xi^{c(t)},\bm\xi^{s(t)}\right),
	\end{equation}

	\item (SP.4) updating auxiliary variables $\bm\xi^c$ and $\bm\xi^s$:
	\begin{equation}
		\max_{\bm\xi^c,\bm\xi^s} \;\tilde{\mathcal G}\left(\bm\xi^{c},\bm\xi^{s}|\mathbf F^{(t)}, \mathbf x^{(t)},\bm\mu^{(t)}\right),
	\end{equation}
\end{itemize}
where $\mathbf v^{(t)}, \mathbf v \in \left\{\mathbf F, \mathbf x, \bm\mu,\bm\xi^c,\bm\xi^s\right\}$ denotes the value of $\mathbf v$ from the last iteration.

\subsection{Updating Transmit Beamforming.}

Focusing on (SP.1), to simplify the objective function, we rewrite it as a quadratic function of $\mathbf F$,
\begin{equation}
	\begin{split}
		\tilde{\mathcal G}(\mathbf F,\mathbf x,\bm\xi^c,\bm\xi^s) = \sum_{k=1}^{K+1}\big(&2\mathcal Re\{\bm\varphi_k^H\mathbf f_k\} - \mathbf f_k^H\mathbf \Lambda_k\mathbf f_k \big)+ B, \\
		&\forall k\in\{1,...,K+1\},
	\end{split}
\end{equation}
where
\begin{gather}
	\begin{split}
		\bm\varphi_k = \varpi_c\sqrt{1+\mu_k}\xi_k^{c*}\mathbf h_k^T\left(\mathbf x\right)+ \varpi_s\sqrt{1+\mu_{K+1}}\\
		\times\alpha_s^*\xi_k^{s*}\mathbf a_s^T\left(\mathbf x\right),
		\text{ for } K \in \{1,...,K\}, 
	\end{split}\\
	\bm\varphi_{K+1} =  \varpi_s\sqrt{1+\mu_{K+1}}\alpha_s^*\xi_{K+1}^{s*}\mathbf a_s^T\left(\mathbf x\right). \\
	\begin{split}
		\mathbf\Lambda_k = \varpi_c\tilde{\mathbf H}_{k}\tilde{\mathbf H}_{k}^H + \varpi_s\left\Vert\bm\xi^{s(t)}\right\Vert^2\Bigg(\sum_{c=1}^C|\alpha_c|^2\mathbf a_{c}\left(\mathbf x\right)\mathbf a_c^H\left(\mathbf x\right), \\
		+ |\alpha_s|^2\mathbf a_s(\mathbf x)\mathbf a_s^H(\mathbf x)\Bigg), k \in\{1,...,K+1\}, 
	\end{split} \\
	\begin{split}
		&B = \varpi_c\sum_{k=1}^{K}\ln(1+\mu_k)+\varpi_s\ln(1+\mu_{K+1}) \\ 
		&- \varpi_c\sum_{k=1}^{K} \mu_k  - \varpi_s\mu_{K+1}-\varpi_c\sum_{k=1}^K|\xi_k^c|^2\sigma_k^2 - \varpi_s\left\Vert\bm\xi^2\right\Vert^2\sigma_s^2.
	\end{split}
\end{gather}
From above derivations, $B$ is a constant for (SP.1), and $\tilde{\mathbf H}_{k} = [\xi^c_1\mathbf h_1, \cdots, \xi^c_K\mathbf h_K]$.

Since $\mathbf\Lambda_k$ is positive definite, (SP.1) is a convex problem that can be efficiently solved by standard solvers such as CVX with high complexity. To reduce the computational complexity of solving this sub-problem, we propose a low-complexity algorithm solving $\mathbf F$ by deriving the closed-form expression based on Lagrangian dual decomposition method \cite{boyd2004convex}. The Lagrangian of the problem can be first derived as
\begin{equation}
	\begin{split}
	\mathcal L(\mathbf F, \lambda) = -\tilde{\mathcal G}\left(\mathbf F|\mathbf x^{(t)},\bm\mu^{(t)},\bm\xi^{c(t)},\bm\xi^{s(t)}\right) \\+ \lambda\left(\tr\left(\mathbf F^H\mathbf F\right) - P_0\right),
	\end{split}
\end{equation}
where $\lambda \ge 0$ is the Lagrangian multiplier corresponding to the power constraint. The KKT conditions are used to solve the dual problem as
\begin{gather}
	\frac{\partial \mathcal L(\mathbf F, \lambda)}{\partial \mathbf F} = \mathbf 0,\label{KKT_stationarity} \\
	\tr\left(\mathbf F^H\mathbf F\right) - P_0 \le 0,\label{KKT_feasibility} \\
	\lambda \ge 0,  \label{KKT_dual_feasibility}\\
	\lambda\left(\tr\left(\mathbf F^H\mathbf F\right) - P_0\right) = 0. \label{KKT_complementary_slackness}
\end{gather}

First, by solving (\ref{KKT_stationarity}), we can obtain the optimal solution of $\mathbf F$ as
\begin{equation}
	\mathbf f_k(\lambda) = \left(\left(\mathbf\Lambda_k^{(t)T} + \lambda\mathbf I\right)^\dagger\right)^*\bm\varphi_k^{(t)}, \;\forall k = \{1,\cdots,K+1\}. \label{update_F}
\end{equation}
The value of $\lambda$ needs to be chosen to satisfy the dual feasibility (\ref{KKT_dual_feasibility}) and the complementary slackness condition (\ref{KKT_complementary_slackness}). If the primal feasibility (\ref{KKT_feasibility}) is satisfied when $\lambda = 0$, the optimal beamforming is $\mathbf f_k(0)$. Otherwise, an appropriate $\lambda$ needs to be decided to satisfy
\begin{equation}
	h(\lambda) = \tr\left(\mathbf F^H(\lambda)\mathbf F(\lambda) \right)- P_0 = 0.
\end{equation}
It can be proved that $h(\lambda)$ is monotonically decreasing with respect to $\lambda$ \cite{JSAC_KKT}, and thus the bisection method can be adopted to find the solution of $\lambda$, which is summarized in \textbf{Algorithm \ref{bisection}}.

\begin{algorithm}[t]  
	\caption{Bisection Method for Searching dual variable $\lambda$.}
	\begin{algorithmic}[1]  
		\State \textbf{Initialize} upper and lower bound $\lambda_\text{max}$, $\lambda_\text{min}$, tolerance $\varepsilon$ and iteration index $l = 0$.
		\Repeat 
		\State Compute $\lambda^{(l)} = (\lambda_\text{min} + \lambda_\text{max})/2$.
		\State Replace $\lambda^{(l)}$ in $\mathbf F(\lambda)$ and compute $h(\lambda^{(l)})$.
		\State If $h(\lambda^{(l)}) > P_0$, set $\lambda_\text{min} = \lambda^{(l)}$. Otherwise, set $\lambda_\text{max} = \lambda^{(l)}$.
		\State  Set iteration index $l = l+1$.
		\Until{$\vert h(\lambda^{(l)}) - P_0\vert \le \varepsilon$.}
		\State \textbf{Output}: optimal dual variable $\lambda^\star$.
	\end{algorithmic} 
	\label{bisection}
\end{algorithm}

\subsection{Updating Antenna Positions.}

With fixed beamformer $\mathbf F$, we update antenna positions $\mathbf x$ by solving (SP.2). Given that the objective function $\tilde{\mathcal G}(\mathbf x|\mathbf F^{(t)},\bm\mu^{(t)},\bm\xi^{c(t)},\bm\xi^{s(t)})$ is neither concave nor convex, but it is differentiable across the entire feasible region, the gradient ascent (GA) method can be utilized to find a local stationary point. However, the moved antennas after GA may violate the feasible region, as well as the minimum distance constraints. One possible solution is to check the satisfaction of the position constraints after each update and proceed until any violation occurs, which is called as DGA algorithm. Nevertheless, due to the non-convexity caused by the complex form and the strict constraints, the DGA scheme might converge to a nearby stationary point or stop at a constraint boundary, resulting in sub-optimal performance. To address this issue, we apply a preliminary search along $\mathcal A$ to find good starting points for the GA operation. Considering that the stop criteria in DGA method may lead to high performance loss, a more effective approach is to initially ignore the constraints during the GA process, and then project the antenna positions that violate the constraints into the nearest feasible point, after the GA operation.

Based on the above discussions, we propose an SPGA based approach to solve (SP.2). The SPGA algorithm for this problem includes three stages: i) initial point search, ii) gradient ascent updating, and iii) feasibility region projection. Here, we use $\tilde{\mathcal G}$ to replace $ \tilde{\mathcal G}(\mathbf x,\mathbf F,\bm\mu,\bm\xi^c,\bm\xi^s)$ for simplicity. The detailed description of the SPGA algorithm for antenna position optimization is as follows.

\subsubsection{Initial point search}

We first set discrete search points $\mathcal X$ on the feasible region $\mathcal A$. $\mathcal X$ is defined as a set of uniform points covering $\mathcal A$. For each antenna $n$, we find an initial $x_n$ that maximizes $\tilde{\mathcal G}$ as
\begin{equation}
	x_n^{(0)} = \arg\min_{x_n \in \mathcal X}\tilde{\mathcal G}, \; n\in\{1,\cdots,N\}. \label{Initial_search}
\end{equation}

\subsubsection{Gradient ascent updating}

With given $x_n^{(0)}$, we update the positions of the $N$ antennas alternatively. The gradient $\nabla_{\mathbf x} \tilde{\mathcal G}$ can be written as
\begin{equation}
	\nabla_{\mathbf x}\tilde{\mathcal G} = \left[\frac{\partial \tilde{\mathcal G}(x_n)}{\partial x_1}, \frac{\partial \tilde{\mathcal G}(x_n)}{\partial x_2}\cdots, \frac{\partial \tilde{\mathcal G}(x_n)}{\partial x_N} \right]^T.
\end{equation}
To make $\tilde{\mathcal G}$ more tractable to derive $\frac{\partial \tilde{\mathcal G}}{\partial x_n}$, $\tilde{\mathcal G}$ can be first rewritten as
\begin{equation}
	\begin{split}
	&\tilde{\mathcal G}(x_n) = \varpi_c\sum_{k=1}^K\sqrt{1+\mu_k} \mathcal F_{1,k}(x_n) \\&-\varpi_c \sum_{k=1}^K|\xi_k^c|^2 \sum_{j=1}^{K+1}\mathcal F_{2,k,j}(x_n) 
	 + \varpi_s\sqrt{1+\mu_{K+1}}\mathcal F_3(x_n) \\&- \varpi_s\Vert\bm\xi^s\Vert^2\sum_{c=1}^C\mathcal F_{4,c}(x_n) - \varpi_s\Vert\bm\xi^s\Vert^2\mathcal F_5(x_n) + \text{Constant},
	\end{split}
\end{equation}
where   
\begin{gather}
	\mathcal F_{1,k}(x_n) = 2\mathcal Re\{\xi_k^c\mathbf h_k^H(x_n)\mathbf f_k\}, \\
	\mathcal F_{2,k,j}(x_n) = \left|\mathbf h_k^H(\mathbf x_n)\mathbf f_j\right|^2, \\
	\mathcal F_3(x_n) = 2\mathcal Re\{\alpha_s\mathbf a_s^H(x_n) \mathbf F\bm\xi^s\}, \\
	\mathcal F_{4,c}(x_n) = \Vert\alpha_c\mathbf a_c^H(x_n)\mathbf F\Vert^2, \\
	\mathcal F_5(x_n) = \Vert\alpha_s\mathbf a_s^H(x_n)\mathbf F\Vert^2.
\end{gather}
Thus, $\frac{\partial \tilde{\mathcal G}(x_n)}{\partial x_n}$ can be derived as
\begin{equation}
	\begin{split}
	&\frac{\partial \tilde{\mathcal G}(x_n)}{\partial x_n} = \varpi_c\sum_{k=1}^K\sqrt{1+\mu_k} \frac{\partial\mathcal F_{1,k}(x_n)}{\partial x_n} -\varpi_c \\
	&\times \sum_{k=1}^K|\xi_k^c|^2 \sum_{j=1,j\neq k}^{K+1} \frac{\partial\mathcal F_{2,k,j}(x_n)}{\partial x_n} 
	+ \varpi_s\sqrt{1+\mu_{K+1}}\frac{\partial\mathcal F_3(x_n)}{\partial x_n} \\
	&-\varpi_s\Vert\bm\xi^s\Vert^2\sum_{c=1}^C\varpi_s\frac{\partial\mathcal F_{4,c}(x_n)}{\partial x_n} - \varpi_s\Vert\bm\xi^s\Vert^2\varpi_s\frac{\partial\mathcal F_5(x_n)}{\partial x_n},
	\end{split}
\end{equation}

The partial derivatives w.r.t. $x_n$ are derived as (\ref{Gra_F1} - \ref{Gra_F4}) at the bottom of the next page, where $f_{k,n}$ denotes the $n$-th element of $\mathbf f_k$, $a_n(\theta) = e^{j\frac{2\pi}{\lambda}x_n\cos\theta}$, $\bar{\bm\xi} = \mathbf F\bm\xi^s=[\bar{\xi}_1, \cdots, \bar{\xi}_N]^T$, $\tilde{F}_{m,n}$ denotes the $(m,n)$-th element of $\tilde{\mathbf F}$.
\begin{figure*}[b]
	\hrule
\begin{gather}
	\frac{\partial\mathcal F_{1,k}(x_n)}{\partial x_n} =2\mathcal Re\Bigg\{-j \xi_k^c\sqrt{\frac{N}{L_p}}\sum_{l=1}^{L_p}\rho_{k,l}^*f_{k,n} 
	\times \frac{2\pi}{\lambda}\cos\theta_l  a_n^*(\theta_{k,l})\Bigg\}, \label{Gra_F1}\\
	\begin{split}
	\frac{\partial \mathcal F_{2,k,j}(x_n)}{\partial x_n} =\frac{N}{L_p}2\mathcal Re\Bigg\{ \sum_{l=1}^{L_p}\sum_{m=1,\atop m\neq n}^N-j\frac{2\pi}{\lambda}\cos\theta_{k,l}|\rho_{k,l}|^2f_{j,n}f_{j,m}^*a^*_n(\theta_{k,l})a_m(\theta_{k,l}) \\
	+ \sum_{l=1}^{L_p}\sum_{p=1,\atop p\neq l}^{L_p}-j\frac{2\pi}{\lambda}(\cos\theta_{k,l}-\cos\theta_{k,p})\rho_{k,l}^*\rho_{k,p} |f_{j,n}|^2 a^*_n(\theta_{k,l}) a_n(\theta_{k,p})  \\
	+ \sum_{l=1}^{L_p}\sum_{p=1,\atop p\neq l}^{L_p}\sum_{m=1,\atop m\neq n}^N -j\frac{2\pi}{\lambda}\cos\theta_{k,l}\rho_{k,l}^*\rho_{k,p} f_{j,n}f_{j,m}^*a^*_n(\theta_{k,l})a_m(\theta_{k,p}) \Bigg\},   
	\end{split}  \label{Gra_F2}\\
	\frac{\partial\mathcal F_3(x_n)}{\partial x_n} = 2\mathcal Re\left\{-j\alpha_s\frac{2\pi}{\lambda}\cos\theta_s\bar{\xi}_n  a_n^*(\theta_s)\right\}, \\
	\frac{\partial \mathcal F_{4,c}(x_n)}{\partial x_n} = |\alpha_c|^2 2\mathcal Re\left\{\sum_{m=1,m\neq n}^Nj\frac{2\pi}{\lambda}\cos\theta_c a_m^*(\theta_c)a_n(\theta_c)\tilde{ F}_{m,n}\right\}, \label{Gra_F4} \\
	\frac{\partial \mathcal F_{5}(x_n)}{\partial x_n} = |\alpha_s|^2 2\mathcal Re\left\{\sum_{m=1,m\neq n}^Nj\frac{2\pi}{\lambda}\cos\theta_s a_m^*(\theta_s)a_n(\theta_s)\tilde{ F}_{m,n}\right\} \text{, where }\tilde{\mathbf F} = \mathbf F\mathbf F^H, \label{Gra_F5}
\end{gather}
\end{figure*}

Hence, with other antennas at fixed positions,   $x_n$ can be alternatively updated in the inner loop as
\begin{equation}
	x_n^{(i+1)} = x_n^{(i)} + \kappa^t\nabla_{x_n}\tilde{\mathcal G},\label{update_u1}, n \in \{1,\cdots, N\},
\end{equation}
where $(i)$ indicates the value obtained from the last iteration in the inner loop for antenna position optimization, $\kappa^t$ denotes the step size for GA in each iteration. This inner updating lasts until the objective converges to a stationary point.

\subsubsection{Feasible region projection}

The updated results may not satisfy the position constraints for the antennas. Therefore, the last step is to project the optimized antenna positions into a a feasible point satisfying the movable region and distance constraints \cite{secureMA}. Note that after optimizing $\mathbf x$, the sequential arrangement of the array elements may be perturbed, i.e. not satisfy $x_1 \le x2 \le \cdots \le x_N$. Hence, different from \cite{secureMA}, we rearrange the indices of antenna elements as $\tilde x_m$ as an one-one mapping, where $X_\text{min} \le \tilde x_1 \le \tilde x_2 \le \cdots \le \tilde x_N \le X_\text{max}$, and each index $m$ has its corresponding $n$. Recalling the constraints (\ref{Cons_ant_region}) and (\ref{Cons_dis_linear}) for linear array, we can obtain
\begin{equation}
	\begin{cases}
		\tilde{x}_1 \ge X_\text{min}, \\ \tilde x_2 - \tilde x_1 \ge D_0,\\  \ \ \ \ \ \vdots \\\tilde x_N - \tilde x_{N-1} \ge D_0, \\ X_\text{max} \ge \tilde x_N.
	\end{cases}
\end{equation}
Then, it is intuitive to determine the projection function to update $\tilde x_n^\star$ one by one as
\begin{equation}
	\red{
	\left\{ \begin{aligned}
	& \tilde{x}_1^{(t+1)} = \max \left(X_{\min}, \min\left(\tilde{x}_1, X_{\max} - (N-1)D_0\right)\right), \\
	& \qquad \vdots \\
	& \tilde x_n^{(t+1)} = \max ( \tilde x_{n-1} + D_0, \min (\tilde x_n, X_{\max} - (N-n)D_0) ), \\
	& \qquad \vdots \\
	& \tilde x_N^{(t+1)} = \max (\tilde x_{N-1} +D_0, \min (\tilde x_N, X_{\max})).
	\end{aligned}\right.} \label{x_projection}
\end{equation}

Finally, we simply assign the values of $x_n^{(t+1)}$ according to the previous one-one mapping from $\tilde x_n^{(t+1)}$.

\subsection{Updating Auxiliary Variable for Lagrangian Dual Transform.}

Regarding (SP.3), we update $\bm\mu$ by taking $\frac{\partial \tilde{\mathcal G}\left(\bm\mu|\mathbf F^{(t)}, \mathbf x^{(t)},\bm\xi^{c(t)},\bm\xi^{s(t)}\right)}{\partial \bm\mu} = 0$, which gives
\begin{equation}
	\mu_k^{(t+1)} =
			\frac{R_k^{(t)2} + R_k^{(t)}\sqrt{R_k^{(t)2}+4}}{2}\; k \in \{1,\cdots,K+1\}, \label{update_mu}
\end{equation}
where $R_k^{(t)} = \mathcal Re\left\{\xi_k^{c(t)}\mathbf h_k^{(t)H}\mathbf f_k^{(t)} \right\},\,k = \{1,\cdots,K\}$, $R_{K+1}^{(t)} = \mathcal Re\left\{\alpha_s\mathbf a_s^{(t)H}\mathbf F^{(t)}\bm\xi^{s(t)} \right\}$.

\subsection{Updating Auxiliary Variable for Quadratic Transform}

Given fixed $\mathbf F$ and $\mathbf x$, we can update the auxiliary variables $\bm\xi^c$ and $\bm\xi^s$ by solving (SP.4). Because $\tilde{G}(\bm\xi^c,\bm\xi^s|\mathbf F^{(t)},\bm\mu^{(t)},\mathbf x^{(t)})$ is a concave function w.r.t. $\bm\xi^c$ and $\bm\xi^s$ without any constraint, the optimal values can be obtained by solving $\frac{\partial \tilde{G}(\bm\xi^c,\bm\xi^s|\mathbf F^{(t)},\bm\mu^{(t)},\mathbf x^{(t)})}{\partial\bm\xi^c} = 0$ and $\frac{\partial \tilde{G}(\bm\xi^c,\bm\xi^s|\mathbf F^{(t)},\bm\mu^{(t)},\mathbf x^{(t)})}{\partial\bm\xi^s} = 0$, which gives
\begin{gather}
	\xi_k^{c(t+1)} = \frac{\left(\mathbf f_k^{(t)}\right)^H\mathbf h_k\left(\mathbf x^{(t)}\right)}{\sum_{j=1}^{K+1}\left\vert\hk^H\left(\mathbf x^{(t)}\right)\left(\mathbf f_j^{(t)}\right)\right\vert^2 +\sigma_k^2},\; k = \{1,\cdots,K\}, \label{update_xic}\\
	\bm\xi^{s(t+1)} = \frac{\alpha_s^*\left(\mathbf F^{(t)}\right)^H\mathbf a_s\left(\mathbf x^{(t)}\right)}{ I_R+ \sigma_s^2}, \label{update_xis}
\end{gather}
where $I_R = \sum_{c=1}^C\left\Vert\alpha_c\mathbf a_{c}^H\left(\mathbf x^{(t)}\right)\mathbf F^{(t)}\right\Vert^2 + \left\Vert\alpha_s\mathbf a_s^H\left(\mathbf x^{(t)}\right)\mathbf F^{(t)}\right\Vert^2$.

\subsection{Overall Algorithm and Complexity Analysis}

Based on the above derivations, the overall algorithm for the MA array is summarized in \textbf{Algorithm \ref{OverallAlg}}.

\begin{algorithm}[t]  
	\caption{Proposed flexible beamforming optimization for MAs.}
	\begin{algorithmic}[1]  
		\State \textbf{Initialize} $\mathbf F^{(0)}$, $\bm\xi^{c(0)}$, $\bm\xi^{s(0)}$. Set iteration index $t = 0$.
		\Repeat \; outerloop
			\State  Update $\mathbf F^{(t+1)}$ by solving (SP.1) as (\ref{update_F}) and \textbf{Algorithm \ref{bisection}}.
			\State  Step i): \red{search for initial starting points $\mathbf x^{(0)}$}.
			\Repeat\; inner loop
				\State  Step ii): alternatively update $x_n^{(i+1)}$ as (\ref{update_u1}).
			\Until {Converges.}
			\State  Step iii): Project the updated $\mathbf x$ into the feasible region to obtain $\mathbf x^{(t+1)}$ \red{as (\ref{x_projection})}.
			\State Update the auxiliary variable $\bm\mu^{(t+1)}$  as (\ref{update_xic}) and (\ref{update_xis}), respectively.
			\State Update the auxiliary variables $\bm\xi^{c(t+1)}$ and $\bm\xi^{c(t+1)}$ as (\ref{update_mu}).
			\State Update iteration index $t = t+1$.
		\Until{The value of the objective function converges.}
		\State \textbf{Output}: optimal $\mathbf F^\star$, $\mathbf x^\star$.
	\end{algorithmic} 
		\label{OverallAlg}
\end{algorithm}

The total complexity of \textbf{Algorithm 2} stems from the solving the four sub-problems. The computational cost of updating $\mathbf F$ is mainly determined by the matrix inversion as $\mathcal O(N^3)$. To update $\mathbf x$, the complexity is dominated by computing the derivative $\frac{\partial \tilde{\mathcal G}(x_n)}{\partial x_n}$, which mainly depends computing $\frac{\partial \mathcal F_{2,k,j}(x_n)}{\partial x_n}$ by (\ref{Gra_F2}). Thus, the complexity for solving (SP.2) can be expressed as $\mathcal O(I_gN^2K^2L_p^2 )$, where and $I_g$ denotes the number of iterations for gradient updating. To update the auxiliary variables as (\ref{update_mu}), (\ref{update_xic}), and (\ref{update_xis}), the complexities are $\mathcal O(NK+K^2)$, $\mathcal O(NK^2)$, and $\mathcal O(NK)$. Hence, the total computational complexity of the proposed algorithm for this flexible beamforming is dominated by GA and can be written as $\mathcal O(I_aI_gN^2K^2L_p^2))$, where $I_a$ denotes the number of iterations of the AO among the sub-problems.

\section{Numerical Results}

In this section, we perform numerical simulations and provide the results to verify the effectiveness and evaluate the performance of our proposed flexible beamforming-ISAC algorithm. 

In the ISAC system, we consider $K=4$ users and $C=3$ clutters. Two cases, $N=4$ and $N=8$ transmit MAs are assumed to evaluate the corresponding performance, respectively. The lower bound of the feasible region for MAs is set as $X_{\min} = 0$, while $X_{\max}$ is a varying parameter to analyze the performance. We set the minimum distance between two adjacent antennas as $D_0 = \frac{\lambda}{2}$. The users and clutters are randomly distributed within the range of $[0,\pi]$, and the target is located at $60^\circ$. $Lp=13$ paths are assigned for each communication user. The complex channel gain, as well as the complex coefficients for target and clutters follow identical CSCG distribution, i.e. $\rho_{k,l}, \alpha_s,\alpha_c \sim \mathcal{CN}(0,1)$. The noise power for each user and target is normalized as $\sigma^2 = \sigma_k^2 = \sigma_s^2 = 1$. \red{The average SNR is assumed as $\text{SNR} = P_0/\sigma^2$}. The wavelength is $\lambda = 0.1$ m. \red{We perform 200 Monte Carlo simulations for the following results.}

In addition to our proposed SPGA- and FP-based flexible beamforming with MAs (\textbf{SPGA-FP, MAs}), we have five baseline schemes to compare the performance improvement of our algorithm, namely i) direct GA- and FP-based flexible beamforming with MAs (\textbf{DGA-FP, MAs}) \cite{MA-MU4}, ii) FP based beamforming with FPAs (\textbf{FP, FPAs}), iii) SPGA-based antenna position optimization and random beamforming with MAs (\textbf{SPGA-RBF, MAs}), iv) DGA-based antenna position optimization and random beamforming with MAs (\textbf{DGA-RBF, MAs}), v) random beamforming with FPAs (\textbf{RBF, FPAs}). \red{The initial setup for the MA Tx array with DGA-based baselines is as a ULA with $\frac{\lambda}{2}$ space between adjacent elements. For SPGA-based schemes, the initial positions are obtained as the initial point search in (\ref{Initial_search}).} For baselines with FPAs, the transmit array is configured as the initial ULA.

\subsection{Overall Utility versus number of iterations.}
\begin{figure}[t]
	\centering
	\includegraphics[width=1\linewidth]{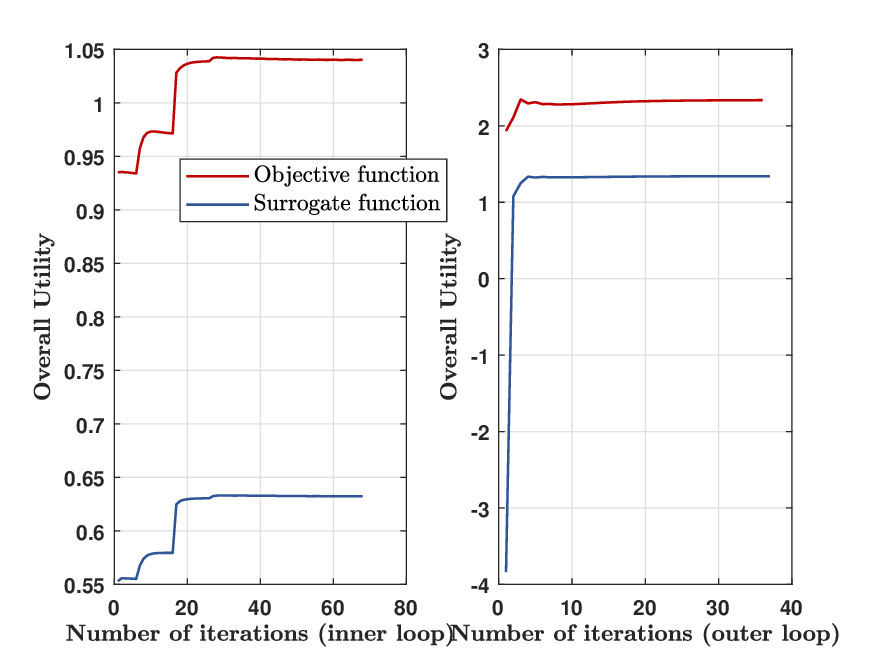}
	\caption{Convergence performance of the inner and outer loops. $N = 4$, \red{$\SNR = 0$ dB}, $X_\text{max} = 10\lambda$, $\varpi_c = 0.5$.}
	\label{Convergence}
\end{figure}
The convergence performance of both the inner and the outer loop of the proposed algorithm can be seen in Fig. \ref{Convergence}, where the both loops shows fast convergence. We compare the original objective function of (P1) with the surrogate function $\tilde{\mathcal G}(\mathbf F,\mathbf x, \bm\mu,\bm\xi^c,\bm\xi^s)$, and the same trend can be observed, demonstrating the effectiveness of the surrogate function in equivalently maximizing the objective function. Interestingly, there are 4 noticeable increases in one inner loop for antenna position optimization, due to the alternating GA update for each antenna.

\subsection{Sum rate and MI versus \red{SNR} and maximum moving region.}

\begin{figure}[t]
	\centering
	\includegraphics[width=0.95\linewidth]{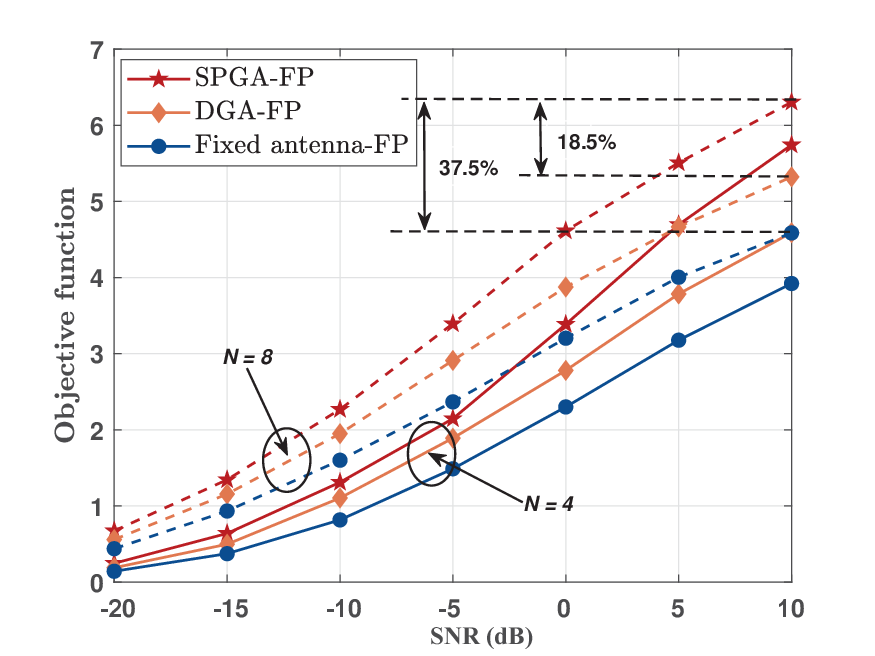}
	\caption{Objective function versus \red{SNR}. $\varpi_c = 0.5$, $X_\text{max} = 10\lambda$.}
	\label{OFvsP0}
\end{figure}

\begin{figure}[t]
	\centering
	\includegraphics[width=0.95\linewidth]{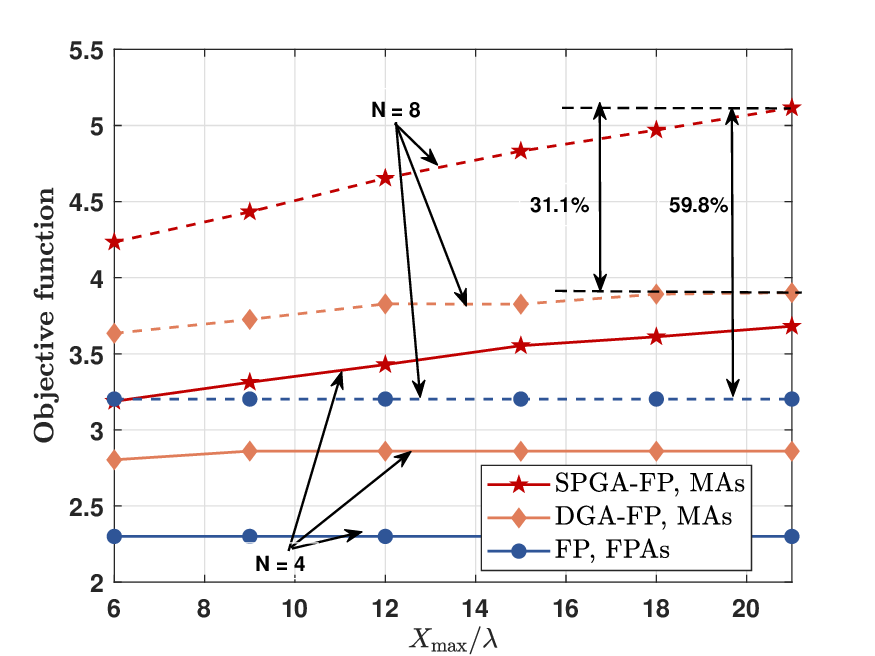}
	\caption{Objective function versus $X_\text{max}$. $\varpi_c = 0.5$, \red{$\SNR = 0$ dB.}}
	\label{OFvsXmax}
\end{figure}

Fig. \ref{OFvsP0} shows how the objective function (the sum rate and MI) \red{varies with transmit SNR}. We set 2 cases, i.e. $N=4$ and $N=8$ MA elements, represented by the solid and the dash lines. With the proposed FP based beamforming design, we compare our proposed SPGA algorithm with DGA and FPAs configuration to evaluate the performance improvement of the proposed antenna position optimization. The objective function grows rapidly with both increasing transmit power and number of transmit antennas. This is because the signal strength can be enhanced by larger transmit power, while the beamforming gain can be achieved by more antennas. Besides, the SPGA algorithm outperforms the other two schemes in antenna position design, \red{especially in high SNR region provided by larger transmit power.} Specifically, with 8 transmit MAs, the SPGA-FP flexible beamforming achieves $18.5\%$ and $37.5\%$ performance gains compared to DGA and FPAs, respectively, through antenna position optimization. \red{It can be seen that 4 transmit MAs with SPGA scheme gradually outperforms 8 FPAs when the transmit SNR grows, due to higher flexibility of beamforming design with MAs.}

\red{Fig. \ref{OFvsXmax}} reveals the objective function versus the maximum moving region. With fixed $X_\text{min} = 0$, $X_\text{max}$ is increasing from $6\lambda$ to $21\lambda$. The results show that the SPGA-FP flexible beamforming significantly improves the ISAC performance, particularly with extended moving region due to more flexible array response reconfiguration. On the contrary, the fixed array has limited performance despite the optimal beamforming design is applied. \red{It is worth noting that utilizing only 4 MAs with the SPGA algorithm demonstrates better performance than 8 FPAs, particularly in scenarios with a large feasible moving region.} MAs with the DGA algorithm has a modest performance gain compared to FPAs, however, only a slight increase can be observed with larger $X_\text{max}$. This is because the DGA method tends to converge quickly to a local stationary point near the initial value without a search stage. Furthermore, the strict antenna distance constrains limit the moving range during GA updating.

\subsection{Trade-off between sensing and communication.}
\begin{figure}[t]
	\centering
	\includegraphics[width=0.95\linewidth]{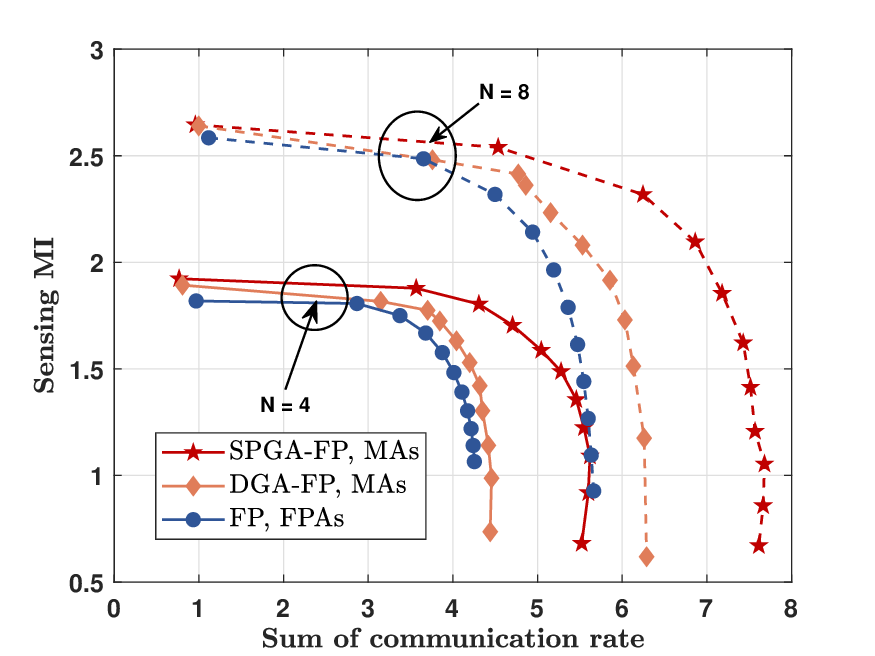}
	\caption{Trade-off between sensing and communication. \red{$\SNR = 0$ dB}, $X_\text{max} = 10\lambda$.}
	\label{RsvsRc}
\end{figure}

We analyze the trade-off between sensing MI and communication rate by adjusting the weighting factors $\varpi_c$ and $\varpi_s$. The MI versus the sum rate is shown in Fig. \ref{RsvsRc}. Dedicated communication function can be implemented by setting $\varpi_c = 1$. Even though the flexible beamforming is designed solely for communication at this scenario, the SR is still capable of receiving weak signals, resulting a low level of sensing MI. Conversely, the same holds true for communication when $\varpi_s = 1$. A larger number of MAs maintain better overall performance from Fig. \ref{RsvsRc} at the cost of higher hardware expense. Hence, appropriate weighting factors and configuration parameters need to be considered in practical engineering.

\subsection{Sensing and communication performance versus \red{SNR}.}

\begin{figure}[t]
	\centering
	\includegraphics[width=0.95\linewidth]{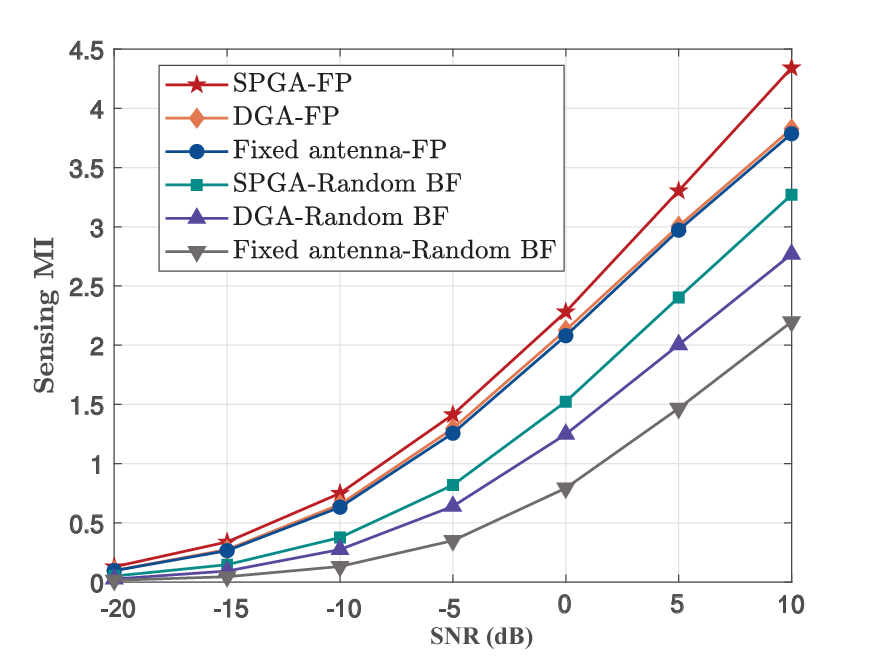}
	\caption{Sensing MI versus \red{SNR}. $N=4$,  $\varpi_s = 0.8$, $X_\text{max} = 10\lambda$.}
	\label{RsvsP0}
\end{figure}

\begin{figure}[t]
	\centering
	\includegraphics[width=0.95\linewidth]{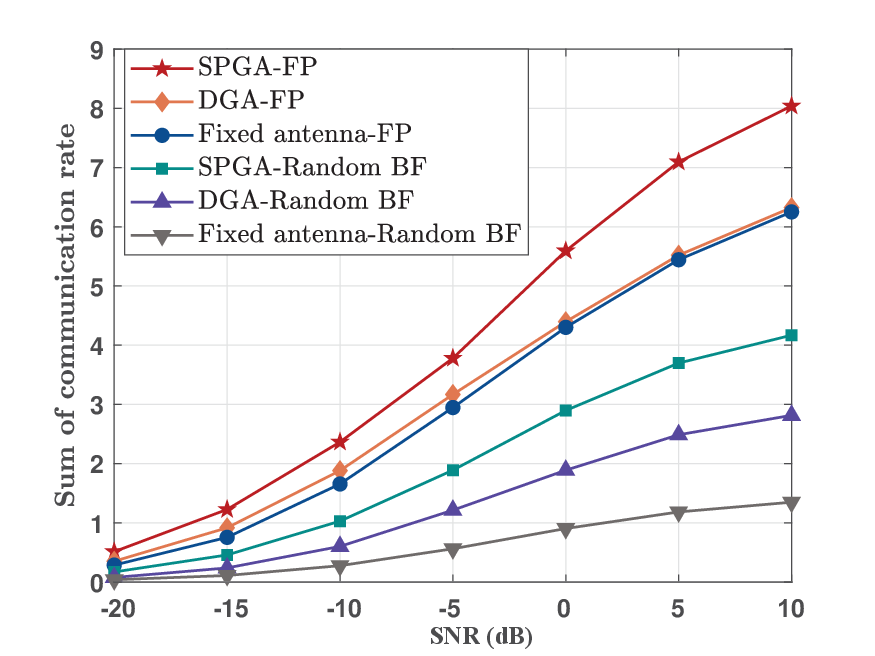}
	\caption{Communication rate versus \red{SNR}. $N=4$, $\varpi_c = 0.8$, $X_\text{max} = 10\lambda$.}
	\label{RcvsP0}
\end{figure}

The sum of sensing and communication performance in the ISAC system has been verified to be efficiently improved above. Then, we analyze the MI and the sum rate, respectively. In Fig. \ref{RsvsP0}, we set $\varpi_s = 0.8$, which means the ISAC system is sensing-focused. It can be observed that sensing MI rises substantially with increasing \red{SNR from -20 dB to 10 dB} . As baselines, DGA and FPAs are utilized to evaluate the performance gain of the SPGA algorithm for antenna position optimization, while random beamforming is adopted to evaluate the performance improvement of the FP-based beamforming design. Consistent with previous results, the proposed SPGA-FP based flexible beamforming noticeably outperforms other baselines in terms of sensing MI. It can be obtained that SPGA-FP based flexible beamforming achieves 32.8\% and 97.6\% performance gains compared to random beamforming with SPGA-optimized MAs and basic FPA array, respectively.

Fig. \ref{RcvsP0} illustrates the sum of communication rate versus \red{SNR} at a communication-focused ISAC with $\varpi_c = 0.8$. Similar trend can be seen as sensing MI in Fig. \ref{RsvsP0}. Interestingly, the performance improvement on communication is more significant than that of sensing. 

\subsection{Pure sensing/communication versus maximum moving region}

\begin{figure}[t]
	\centering
	\includegraphics[width=0.95\linewidth]{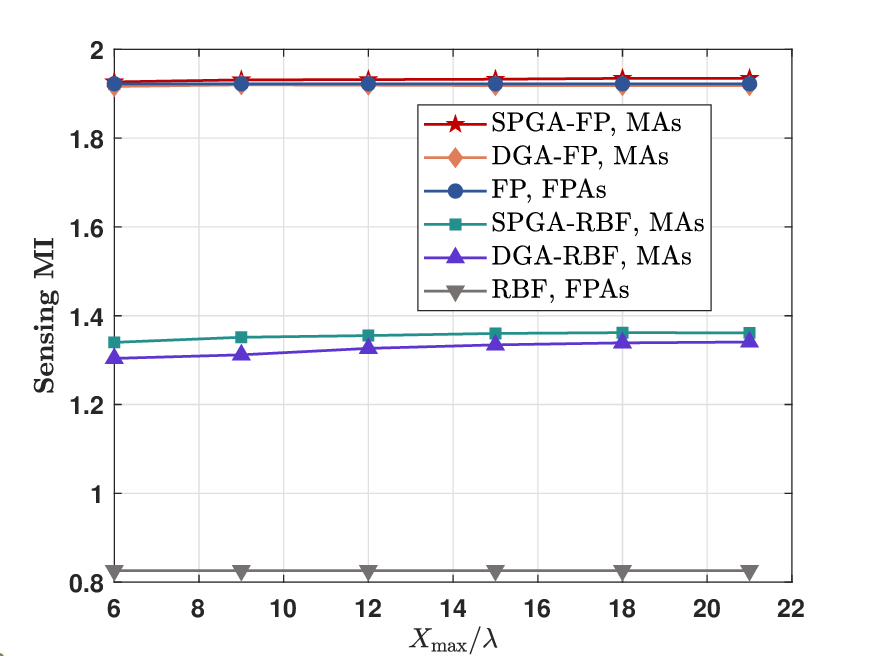}
	\caption{Sensing MI versus $X_\text{max}$. $N=4$,  $\varpi_s = 1.0$, \red{$\SNR = 0$ dB}.}
	\label{MIvsXmax}
\end{figure}

In Fig. \ref{MIvsXmax}, we analyze the sensing MI versus the size of the feasible moving region in pure sensing mode, setting $\varpi_s = 1.0$. It can be seen that the sensing MI almost remains nearly constant as $X_\text{max}$ increases. Although the MA array shows performance improvement without beamforming design, it provides only a marginal performance improvement in sensing MI  compared to FPA array, with the optimized FP-based beamforming design. Additionally, antenna coefficient optimization for beamforming demonstrates higher potential for sensing performance improvement compared to antenna position optimization. The reason behind these results is that only a single path is considered for sensing to receive the echo signal, and the designed beamforming is already sufficient to provide optimal beam gain as well as clutter suppression in sensing. 

\begin{figure}[t]
	\centering
	\includegraphics[width=0.95\linewidth]{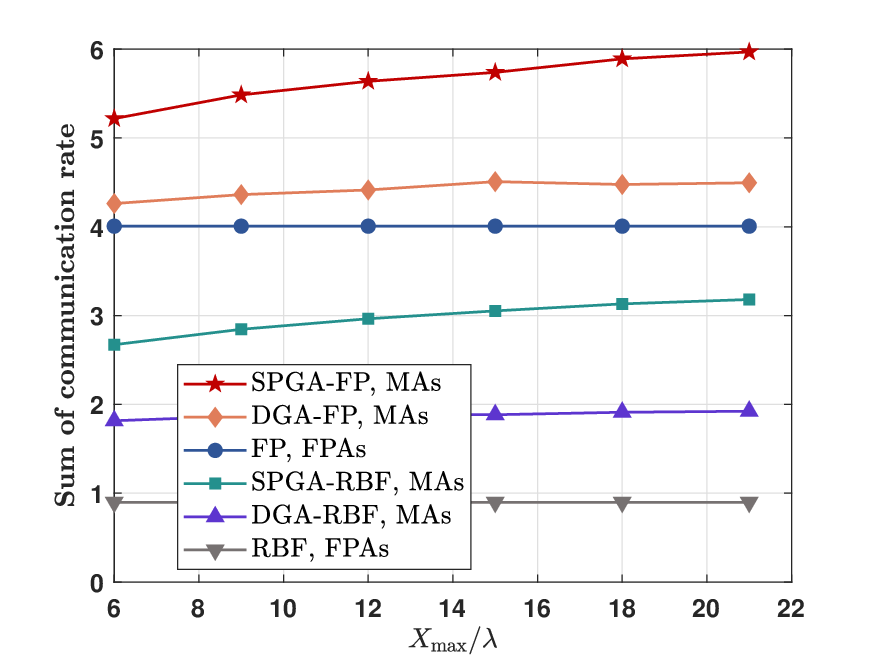}
	\caption{Communication rate versus $X_\text{max}$. $N=4$, $\varpi_c = 0$, \red{$\SNR = 0$ dB}.}
	\label{RcvsXmax}
\end{figure}

In Fig. \ref{RcvsXmax}, pure communication mode is used to analyze how the communication rate varies with an increasing moving region. It can be observed that both antenna coefficient and antenna position design provide significant performance gains in terms of user rate, although antenna coefficient optimization still yields better results. Moreover, with a larger moving region, the proposed SPGA-based antenna position optimization enables an increase in user rate. This demonstrates that MAs are beneficial for multi-path scenarios, and is able to assist beamforming design by adjusting the field response.

\subsection{Beampattern versus angle.}

\begin{figure}[t]
	\centering
	\includegraphics[width=0.95\linewidth]{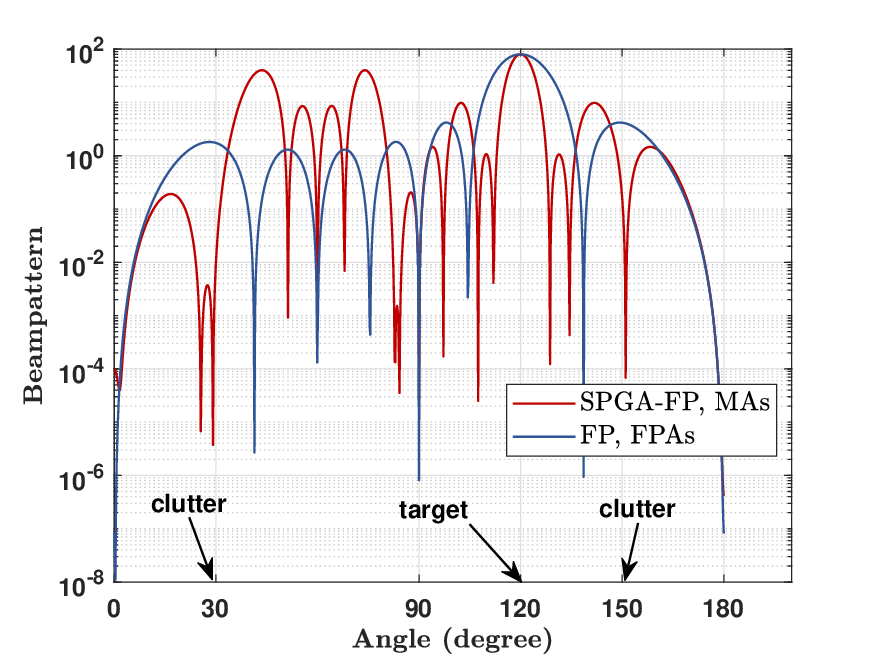}
	\caption{Sensing beampattern versus angle. $N=8$, $\varpi_s = 1.0$, \red{$\SNR = 10$ dB}, $X_\text{max} = 10\lambda$.}
	\label{Beampattern}
\end{figure}

Various metrics are investigated and proved to be effective to measure the sensing performance. We consider the optimized beampattern to intuitively observe how MAs contribute to our ISAC system in sensing function. The beampattern at angle $\theta$ is computed as
\begin{equation}
	\text{BP}(\theta) = \left\Vert\mathbf a^H(\theta)\mathbf F\right\Vert^2.
\end{equation}

To eliminate the interference from communication users, we set $\varpi_s = 1$ to study the sensing-only scenario. $N=8$ MAs and FPAs are configured, respectively. The target is assumed to be located at 120$^\circ$ while 2 clutters are located at 30$^\circ$ and 150$^\circ$, respectively. For FPA array, the target is illuminated by the mainlobe, while the clutter suppression is unsatisfied. Fortunately, we can observe that MAs with SPGA scheme achieves a remarkable clutter suppression effect, thereby improving the SCNR and sensing MI.

\subsection{CRB versus angle.}

\begin{figure}[t]
	\centering
	\includegraphics[width=0.95\linewidth]{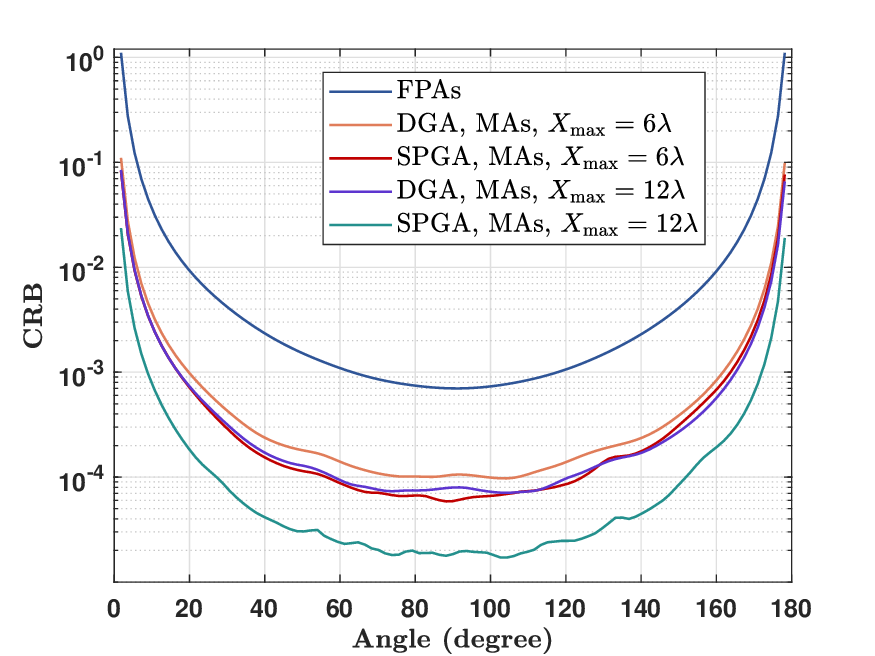}
	\caption{CRB versus angle. $\varpi_s = 0.8$, \red{$\SNR = 0$ dB}.}
	\label{CRB}
\end{figure}

To assess the accuracy of estimating target parameters, we derive the CRB for the target angle in the Appendix, and analyze the effect of the MAs on the CRB. We set random directions of the target and clutters. To focus on comparison between MAs and FPAs, we set the beamforming matrix as a unitary matrix with scale on transmit power. Fig. \ref{CRB} illustrates the variation in CRB values across a 180$^\circ$ range. Due to the asymmetrical response of non-uniform arrays, fluctuations are observed in MA configurations. It is evident from the figure that optimizing antenna positions significantly reduces the CRB for estimating the angle of the target. Moreover, a larger moving region with the proposed SPGA-based algorithm further improves the CRB performance. Hence, maximizing sensing MI is verified to be an effective alternation of minimizing CRB by antenna position optimization, which is intractable due to the complex expression of CRB.

\section{Conclusion}

Overall, this paper studied an MA-enabled ISAC system. Flexible beamforming was investigated by jointly optimizing the antenna coefficients and antenna positions. The goal was to maximize the sum of the communication rate and sensing MI, with a weighting factor to control the priority. To address the non-convexity of the problem, we transformed the objective function using the FP method, and alternately solved four sub-problems. We derived the closed-form expression for updating the beamforming matrix based on the KKT conditions. The positions of the antennas were optimized through the proposed 3-stage SPGA-based scheme. Numerical results confirmed the effectiveness of flexible beamforming with MAs in improving ISAC system performance through the proposed algorithm. Notably, the SPGA-based scheme demonstrated significant performance gains with a large feasible moving region and in high SNR settings, compared to the DGA-based method. Remarkably, the ISAC system with only 4 MAs outperformed the system with 8 FPAs, offering valuable insights for reducing hardware costs in engineering applications. Nevertheless, we observed that MAs performed better in multi-path communication function compared to single-path sensing. In addition to sensing MI, the beampattern and CRB performance for sensing were derived and analyzed. The results showed that MAs effectively suppressed interference suppression, as demonstrated by the beampattern diagram. Furthermore, the proposed MI maximization algorithm proved effective in enhancing sensing accuracy by reducing the estimation CRB. \red{ Additionally, multiple sensing targets are common in practical scenarios. However, more sensing tasks will introduce additional interference, affecting both communication and sensing performance. Therefore, future work could explore advanced techniques for interference suppression to further improve the system performance.}

\appendices

\section{Derivation of CRB}

By sending $T$ sensing symbols, the echo signal over $T$ time indices at the sensing Rx is
\begin{equation}
	\begin{split}
		\mathbf y^H &= \alpha_s\mathbf a_s^H\mathbf F\mathbf S + \sum_{c=1}^C\alpha_c\mathbf a_c^H\mathbf F\mathbf S + \mathbf n^H , \\
		&= \mathbf p_s^H + \sum_{c=1}^C\mathbf p_c^H + \mathbf n^H \in \mathbb C^{1\times T},
	\end{split}
\end{equation}
where $\mathbf S = [\mathbf s(1),\cdots,\mathbf s(T)] \in \mathbb C^{(K+1)\times T}$, $\mathbf n = [n(1),\cdots,n(T)]^T\in \mathbb C^{T\times 1}$, $\mathbf a_s$ and $\mathbf a_c$ denote $\mathbf a_s(\mathbf x)$ and $\mathbf a_c(\mathbf x)$ for simplification.

To analyze the CRB for estimating the target parameters, we regard the target echo signal $\mathbf p_s^H$ as deterministic signal, while the clutter echo $\mathbf p_c^H$ can be viewed as zero-mean Gaussian noise \cite{5672411}. The covariance matrix of $\mathbf p_c^H$ is $\mathbb E\left\{|\alpha_c|^2\mathbf S^H\mathbf F^H\mathbf a_c\mathbf a_c^H\mathbf{FS}\right\} = |\alpha_c|^2\mathbf a_c^H\mathbf F\mathbf F^H\mathbf a_c\mathbf I$. The noise at each time index is independently identically distributed and thus the noise covariance matrix is $\mathbb E\left\{\mathbf n\mathbf n^H\right\} = \sigma_s^2\mathbf I$. Based on the above derivations, the clutter-plus-noise can be viewed as effective noise following $\tilde{\mathbf n} \sim \mathcal{CN}(\mathbf 0,\tilde\sigma_s^2\mathbf I)$, where $\tilde \sigma_s^2 = \sum_{c=1}^C|\alpha_c|^2\mathbf a_c^H\mathbf F\mathbf F^H\mathbf a_c + \sigma_s^2$.

The estimated parameters are represented as $\bm\zeta = [\theta_s, \mathcal Re\{\alpha_s\}, \mathcal Im\{\alpha_s\}]^T$. So the Fisher information matrix (FIM) can be written as \cite{1703855}
\begin{equation}
	\mathbf J = \begin{bmatrix}
		\mathbf J_{\theta\theta} & \mathbf J_{\theta\alpha_R} & \mathbf J_{\theta\alpha_I} \\
		\mathbf J_{\alpha_R\theta} & \mathbf J_{\alpha_R\alpha_R} & \mathbf J_{\alpha_R\alpha_I} \\
		\mathbf J_{\alpha_I\theta} & \mathbf J_{\alpha_I\alpha_R} & \mathbf J_{\alpha_I\alpha_I}
	\end{bmatrix},
\end{equation}
where each element can be derived as
\begin{equation}
	\begin{split}
		\mathbf J_{p,q} &= \text{tr}\left\{\mathbf R_{\tilde{\mathbf n}}^{-1}\frac{\partial \mathbf R_{\tilde{\mathbf n}}}{\partial \zeta_p}\mathbf R_{\tilde{\mathbf n}}^{-1}\frac{\partial \mathbf R_{\tilde{\mathbf n}}}{\partial\zeta_q}\right\} + 2\mathcal Re\left\{\frac{\partial\mathbf p_s^H}{\partial\zeta_p}\mathbf R_{\tilde{\mathbf n}}^{-1}\frac{\partial \mathbf p}{\partial\zeta_q}\right\},  \\
		&= \frac{2}{\tilde{\sigma}_s^2} \mathcal Re\left\{\frac{\partial\mathbf p_s^H}{\partial\zeta_p}\frac{\partial \mathbf p}{\partial\zeta_q}\right\},\; p,q = \{1,2,3\},
	\end{split}
\end{equation}
which can be derived in detail as
\begin{gather}
	\mathbf J_{\theta\theta} = \frac{2T|\alpha_s|^2}{\tilde\sigma^2}\dot{\mathbf a}_s^H\mathbf{FF}^H\dot{\mathbf a_s}, \\
	\mathbf J_{\alpha_R\alpha_R} = \mathbf J_{\alpha_I\alpha_I} = \frac{2T}{\tilde{\sigma}^2} \mathbf a_s^H\mathbf F\mathbf F^H\mathbf a_s, \\
	\mathbf J_{\theta\alpha_R} = \mathbf J_{\alpha_R\theta}^* = \frac{2T}{\tilde \sigma^2} \mathcal Re\left\{ \alpha_s\dot{\mathbf a}_s^H\mathbf F\mathbf F^H\mathbf a_s \right\}, \\
	\mathbf J_{\theta\alpha_I} = \mathbf J_{\alpha_I\theta}^*= \frac{2T}{\tilde\sigma^2}\mathcal Re\left\{ -j\alpha_s\dot{\mathbf a}_s^H\mathbf F\mathbf F^H\mathbf a_s \right\},\\
	\mathbf J_{\alpha_R\alpha_I} = \mathbf J_{\alpha_I\alpha_R} = 0,
\end{gather}
with $\dot{\mathbf a}_s$ denoting $\frac{\partial \mathbf a_s}{\partial \mathbf\theta_s}$, and the $n$-th element can be derived as
\begin{equation}
	[\dot{\mathbf a}_s]_n = -j\frac{2\pi}{\lambda}x_n\sin\theta_s e^{j\frac{2\pi}{\lambda}x_n\cos\theta_s}.
\end{equation}

For target location estimation, we only care the target angle. Therefore, the interested CRB can be computed as
\begin{equation}
	\text{CRB}(\theta_s) = [\mathbf J^{-1}]_{1,1}.
\end{equation}

\ifCLASSOPTIONcaptionsoff
\newpage
\fi

\bibliographystyle{IEEEtran}
\bibliography{Ref}

\end{document}